\documentclass[12pt,preprint]{aastex}
\begin{document}
\title{Evidence for New Unidentified TeV $\gamma$-ray Sources from 
Angularly-Correlated Hot-Spots Observed by  \\
Independent TeV $\gamma$-ray Sky Surveys}
\author{G. Walker, R. Atkins, and D. Kieda}
\affil{High Energy Astrophysics Institute, University of Utah, Salt Lake City, UT 84112}
\email{walker@physics.utah.edu}
\email{ratkins@cosmic.utah.edu}
\email{kieda@physics.utah.edu}
\begin{abstract}
We have  examined the directional 
cross-correlation of statistical `hot-spots' between a Northern Sky 
TeV Gamma Ray Survey by the Milagro Observatory and a similar survey 
by the Tibet Array. We find the directions of these hot-spots 
are angularly uncorrelated between the two surveys for large angular separations 
($\Delta\theta > 4^\circ$), but there appears to be a statistically significant correlation
between hot-spot directions for $\Delta\theta < 1.5^\circ$.
Independent simulations indicate the chance 
probability for the occurrence of this correlation is approximately 
 $10^{-4}$, implying the existence of one or more 
previously  unobserved TeV $\gamma$-ray sources 
in these directions. The data sets are consistent with 
 both point-like sources or diffuse sources with extent of $1^\circ-2^\circ$. 
\end{abstract}


\section{Motivation}
The Milagro observatory and the Tibet Air Shower array are wide field of view 
TeV $\gamma$-ray (1 TeV = 10$^{12}$ eV) observatories
that are capable of monitoring the northern hemisphere  sky on both long and short 
timescales. The Tibet and Milagro detectors
have similar exposures and angular resolutions ($ \leq 1^\circ$) as verified 
by moon shadow analysis \citep{frank_moon,tibet_moon}.  Based on the moon shadow 
analysis Tibet reports a systematic pointing error of 0.1$^{\circ}$ while Milagro reports an 
overall angular resolution of 0.75$^{\circ}$ including pointing errors.
Recent Tibet \citep{tibet_all_sky_2001, tibet_all_sky_2003} 
and  Milagro \citep{milagro_all_sky} northern-hemisphere sky surveys 
have detected  statistical `hot-spots' where excessive numbers of cosmic-rays ($>4 \sigma$ above
expected background level) appear to be concentrated from specific directions. 
Two of these hot-spots are identified with  well known TeV 
sources \citep{milagro_crab,tibet_crab,tibet_421}.
In each sky survey, the remaining hot-spots are consistent with
random statistical fluctuations in the cosmic ray background rate in each direction.
However, if real TeV $\gamma$-ray sources exist with fluxes just below the sensitivity of these  
observatories, then one may expect to see angular 
correlation between the directions of the Milagro sky-survey
hot-spots and the Tibet survey hot-spots, 
with an angular correlation distance equal to a convolution of the angular resolution functions of the
   two detectors.  This may be complicated by pointing errors for weak 
point sources and detector systematics.  Furthermore, it is unclear what angular correlation to 
expect for a diffuse TeV $\gamma$-ray emission region.

\section{Milagro and Tibet All-Sky Analysis}
Both Milagro and Tibet performed a $\gamma$-ray sky survey by plotting 
the angular distribution of reconstructed directions of 
cosmic-rays and $\gamma$-rays on an all-sky map. The sky map is 
divided into finite size angular bins, and  hot-spots in the sky map
are identified  where  a statistically significant
number of excess cosmic-rays and $\gamma$-rays  (above
an average background level)  appear in the selected angular bin.

The Tibet analyses \citep{tibet_all_sky_2001} determine the background 
($\mathrm{N_{off}}$) by the equi-declination method.  This method assumes that the 
background in the same declination band as  the source constitutes a smooth 
background in RA.  For both Tibet sky surveys, the estimated background in 
the signal bin is 
determined by performing a second order $\chi^{2}$ fit to the off source
bins.

The Milagro analysis uses the method of direct integration to estimate the 
background\citep{milagro_crab,morales_2002,Alex_1993}.  The direct integration method works on 
the assumption that cosmic rays create an isotropic background and that the 
acceptance of the detector is independent of trigger rate over some time period (2 hours in the Milagro analysis).  The 
expected number of background events N$_{exp}$ is estimated using 
\begin{equation}
N_{exp}[RA, \delta] = \int \int E(HA,\delta)R(t)\epsilon(HA,RA,t)dtd\Omega.
\end{equation}
The {\it E(HA,$\delta$)} term is the acceptance of the detector in local 
coordinates (HA and declination), {\it R(t)} is the trigger rate 
over some time window (in the case of \citep{milagro_all_sky} the 
window is two hours), and $\epsilon(HA,RA,t)$ is a mapping function 
between local coordinates and celestial coordinates as a function of time. 

The statistical significance $S$ in each angular bin
is  calculated differently for both surveys.  The Milagro 
survey used the method of Li \& Ma (1983). 
The Tibet analyses  calculated  the 
statistical significance of each bin  using  a somewhat simpler technique\citep{tibet_all_sky_2001}. 

The Tibet 2001 sky survey analysis \citep{tibet_all_sky_2001} finds 18 hot-spots (above 4$\sigma$)
which are un-associated with any known TeV $\gamma$-ray source. The Tibet 2003 sky survey \citep{tibet_all_sky_2003} 
find 21 hot spots which are un-associated with known TeV $\gamma$-ray sources, 
but only report the directions of three of these hot-spots in their paper.
In each Tibet 
survey a different non-overlapping data set was used.  Thus the two Tibet surveys should be 
independent of each other.
The  Milagro analysis \citep{milagro_all_sky} reports the directions of 
9 unidentified hot-spots.   Table 1 summarizes the relevant information 
regarding the three surveys.

\section{Angular Correlations  Between Milagro Hot-Spots and Tibet Hot-Spots}
Since the Tibet 2003 analysis only reports an 
incomplete list of hot-spot directions in their sky survey, we 
have limited our analysis to angular correlations between 
the 18 Tibet 2001 hot-spot directions 
and  the 9 Milagro hot-spot directions.  
We compile the measured angular correlation distribution between the two surveys by pairing 
each Milagro hot spot direction with every Tibet 2001 
direction and calculating the angular separation between the pair. We populate 
a histogram with angular differences derived for each possible pair combination between the two surveys. 
Figure 1 illustrates the
resulting histogram distribution of angular differences between the 
two independent sky  survey hot-spot populations. In this plot we have binned 
the data in $2^\circ$ bins, larger than the expected combined angular correlation distance ($1.5^\circ$).

The expected angular correlation distribution for uncorrelated pairs is influenced mostly
by geometrical considerations of field of view of the two instruments, and specifically the
number of possible angular combinations available when random shower directions are seeded over
the fields of view of each instrument. In order to simulate this, we populated 0.1$^\circ \times 0.1^\circ$ sized
bins in right ascension(RA) and declination (Dec) with a sample of events drawn
from a mean background population.  The background population was uniform in RA and followed 
a $\cos(declination - latitude)$ dependence in declination.    
(We also looked at a $\cos^{2}(declination - latitude)$ and a $\cos^{8}(declination - latitude)$ distribution and 
found our results to be very similar.)  Here $latitude$ 
is the specific latitude for each observatory, and $declination$ reflects the range of declination
field of view of each observatory.  In general the distribution of excesses in the sky should be independent of the 
region of the sky (assuming the significance is calculated correctly).  Once an independent simulated sky map was 
generated for each observatory, in accordance with its specific 
latitude and field of view, each sky map was 
binned in a manner  appropriate  to the method employed by
each analysis (a circle for Tibet 2001 
and a square for Milagro).  The background for both simulated 
sky maps were found by averaging 20 bins 
at the same declination, and the statistical significance of each bin population
 was then calculated using the Li \& Ma method for the Milagro simulation, and the Tibet method for the Tibet simulation. 
The Tibet method, as quoted, is 
\begin{equation}
S_{\sigma} = \frac{N_{on} - N_{off}/m} {\sqrt{N_{off}}/m}.
\end{equation}
Where $\mathrm{S_{\sigma}}$ is the significance, $\mathrm{N_{on}}$ is the number of counts in the source bin, 
$\mathrm{N_{off}}$ is the number of counts in the off source bins, and m is the ratio of exposures to the on source 
region and the off source region \citep{tibet_all_sky_2001}.

The simulations for Tibet 2001 produced on average 11  
hot-spots with statistical significance $> 4\sigma$, in good agreement with the observed number.
The simulations for Milagro produced an average of 10 hot spots of similar significance, also
in good agreement with the reported number.
The expected angular correlation distribution for uncorrelated pairs was then compiled
by pairing each simulated Milagro hot-spot with every simulated Tibet hot-spot and
calculating the angular separation between the pair, in a manner identical to that applied 
to the real data (see figure 1).

For large angular separations ($\Delta\theta > 4^\circ$) the measured and simulated 
correlation distributions are in reasonable agreement. At small angular separations
($\Delta\theta < 2^\circ$), there is a statistically significant deviation from
the expected angular correlation distribution for uncorrelated pairs. Three correlated
pairs are found, whereas approximately 0.1 are expected. Each of these pairs is found to have angular 
separation $\leq 1.5^\circ$ between the correlated hot-spots, consistent with expectations from the combined 
angular resolution between the two detectors.  Figure 2 shows the integral Poisson probability for 
finding the observed number of correlations, given the mean value from the simulation.  

The probability for finding 3 hot-spot pairs (within 1.5$^{\circ}$) between the two surveys
 can be estimated by placing the 18 Tibet 2001 locations and the 9 Milagro 
locations randomly and uniformly across the sky in the declination regions used 
in each sky survey. These simulated distributions 
are then searched for coincident hot-spots and the probability of 
 having $N$ hot-spot correlations with $\Delta\theta <1.5^{\circ}$ is compiled
from the fraction of simulations which yield $N$  correlated hot-spot pairs.
(Method 1). This is a reasonable approximation because the distribution of 
hot-spots is found to be relatively uniform across
the observatory's field of view in both 
measured sky survey distributions as well as the above uncorrelated pair 
 angular correlation distribution simulations.

The more extensive angular correlation distribution simulations can also be used to 
independently calculate the probability of observing $N$ hot-spot correlations with 
$\Delta\theta <1.5^{\circ}$ from the fraction of simulations which yield $N$  correlated hot-spot pairs.
(Method 2). The results of our these calculations
 for both methods are presented in Table 3.  The calculations of both
methods are consistent with each other  and indicate that the chance probability 
of finding 3 uncorrelated hot-spot pairs (within 1.5$^{\circ}$) between the two surveys
is small. 

In any analysis of this type, the number of trials must be taken into account.  The Monte Carlo 
simulation method accounts for all  trials except for that associated with the choice of 
a correlation distance of 1.5$^{\circ}$.  In this work our choice of 1.5$^{\circ}$ is based 
upon the expected independently combined angular resolution of Tibet and Milagro 
($\mathrm{\sigma_{comb}=\sqrt{\sigma_{Milagro}^{2} + \sigma_{Tibet}^{2}} \sim 1.5}$).  We did not 
examine correlations on different length scales, but it is important to note from figure 1
 that this result is relatively independent of any reasonable choice of the correlation
distance between $1.5^\circ$ and 4$^\circ$. This would indicate a trials factor
for the angular correlation distance of order of magnitude 1.

However, even if one conservatively assumed trials factor of order 10, the observed deviation from the 
expected random behavior at small angular separations is still statistically compelling.

\section{Results and Discussion}
The coordinates of the three angularly correlated 
hot-spot pairs derived from the Tibet 2001 and Milagro sky surveys 
are given in Table 2. Of the hot-spot pairs, we find Pair A (hot-spots 1 and 5) 
and Pair B (hot-spots 2 and 6) to be the most interesting. Pair A lies on the galactic plane.  
The chance probability of this single pair is 5.4\% using Method 1.
Although this chance probability is marginally interesting, there
 also exists a Tibet 2003 hot-spot of $4.0\sigma$ excess in this region.  
The Tibet 2003 hot-spot is 1.8$^{\circ}$ from the Tibet 2001 hot-spot 
and 3.1$^{\circ}$ from the Milagro hot-spot.  Summing the probabilities for all 
permutations of these three hot-spots, we estimate an overall chance probability of 1.5\% for such a coincidence.  
TeV observations in the direction of Pair A
have been made by the Whipple Collaboration in 1999 (7.2 hours on J2020, which 
is 1$^{\circ}$ south of hot-spot 5) and in 2002 (4.2 hours on hot-spot 5)\citep{walker_whipple}.  
These observations yielded no point-sources of $\mathrm{>200 GeV}$ $\gamma$-rays
at the 0.5 Crab level flux, assuming a  Crab-like power-law energy spectrum.   

The second hot-spot pair correlation (Pair B, hot-spots 2 and 6 in Table 1)
 has a 0.6\% chance of random occurrence (with an angular separation $<0.6^\circ$, 
using Method 1) and is near an X-ray bright
 region of the Cygnus Loop, in the Galactic Plane.  
The third hot-spot pair correlation (Pair C , hot-spots 3 and 7 in Table 2) lies in the same field as Pegasus and 
consists of numerous faint galaxies, but is off the Galactic Plane. 
The Whipple Observatory has not had any contemporaneous 
observations in either of these directions.

\section{Conclusions}
While the hot-spot regions reported
 by the Milagro and the Tibet groups are not statistically significant 
on their own, angular correlations between hot-spots in 
the two sky surveys strongly indicate the possible presence of one or more new, 
unidentified TeV $\gamma$-ray sources 
with $\gamma$-ray flux  just at or slightly below 
the flux sensitivity of each experiment.  

Based on the published upper limits for the Milagro hot-spots the expected flux from these 
possible observations  must be $\sim$ 0.8 times the flux from the Crab Nebula in the 
TeV range in order to have caused these 
fluctuations, and simultaneously avoided strong direct-detections by the two northern-sky surveys.  
The energy spectrum could be a power law.  It is also possible that spectrum is non-conventional.  However there is 
no evidence to suggest either.

It may be fruitful for more sensitive GeV/TeV $\gamma$-ray 
instruments to perform  observations around these source regions to 
search for possible new sources of GeV/TeV $\gamma$-rays. 
However, the sources in question may exhibit 
variability or may be diffuse sources, causing difficulties with
IACT confirmation. Consequently, we suggest that correlated angular analysis 
between all-sky surveys in other wavelengths (such as MeV/GeV Satellite measurements and the AMANDA/ICECUBE 
neutrino detectors ) may
provide additional evidence for new astrophysical sources whose
emission rate falls just slightly below the sensitivity of these  instruments.
\section{Acknowledgments}
We gratefully acknowledge support for this work from the University of Utah
and the National Science Foundation under NSF Grants \#PHY 0079704 and \#PHY 0099580. 
We thank Paul Sommers for useful comments and discussion on this article.  
Lastly we would like to thank the referee for his/her useful comments that 
have improved our paper.

\clearpage

\begin{figure}
\includegraphics[angle=0.,scale = .75]{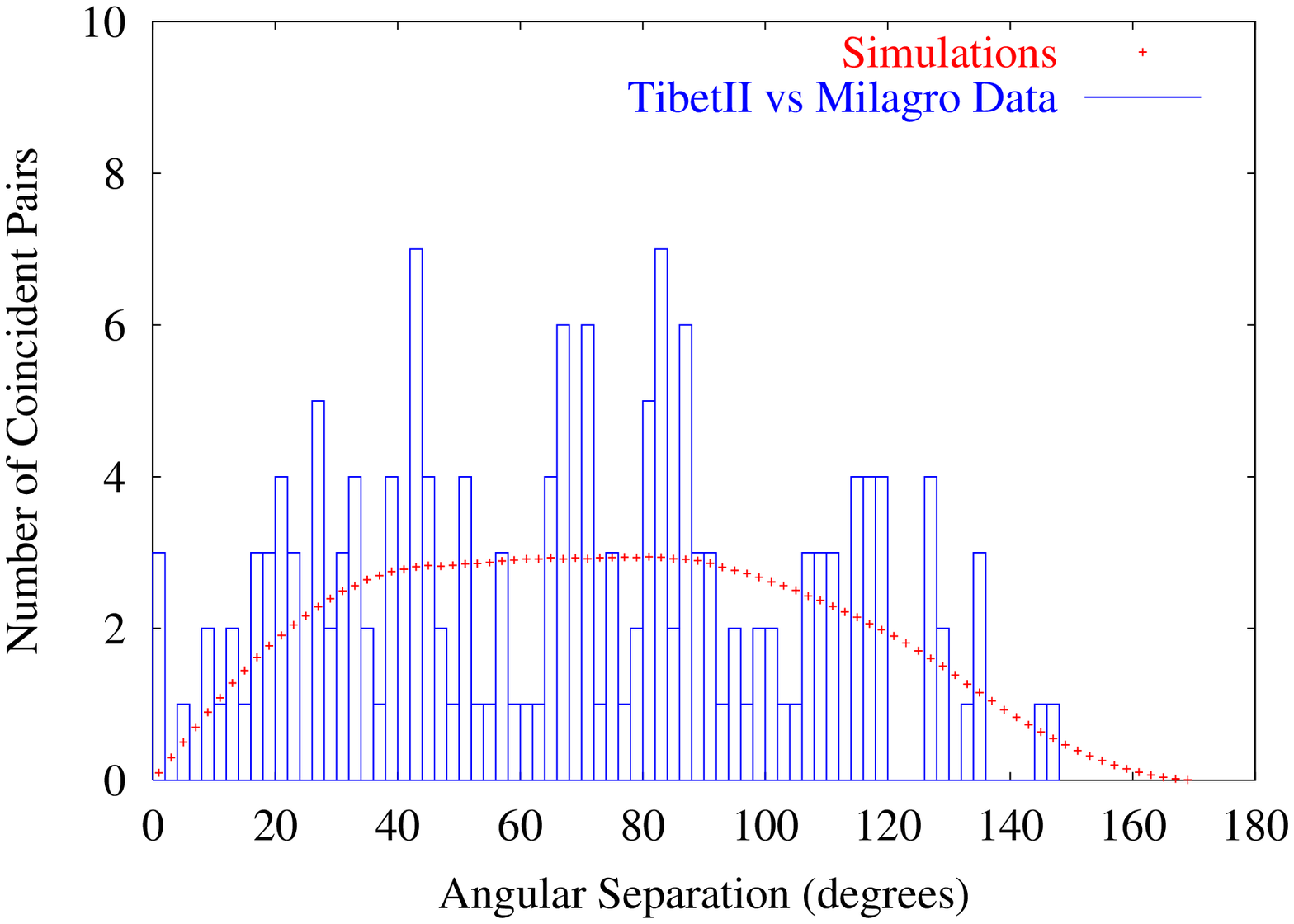}

\caption{Angular correlation distribution compiled from
angular distance from each Tibet hot-spot direction to every
Milagro hot-spot direction. The excess 
number of pairs at small values of angular separation indicates the 
likely presence of one or more new unidentified TeV $\gamma$ -ray sources.  The uncertainty in the 
simulated data points is just the Poisson uncertainty (square root of the mean).}
\end{figure}
\clearpage
\begin{figure}
\includegraphics[angle = 0., scale = 0.75]{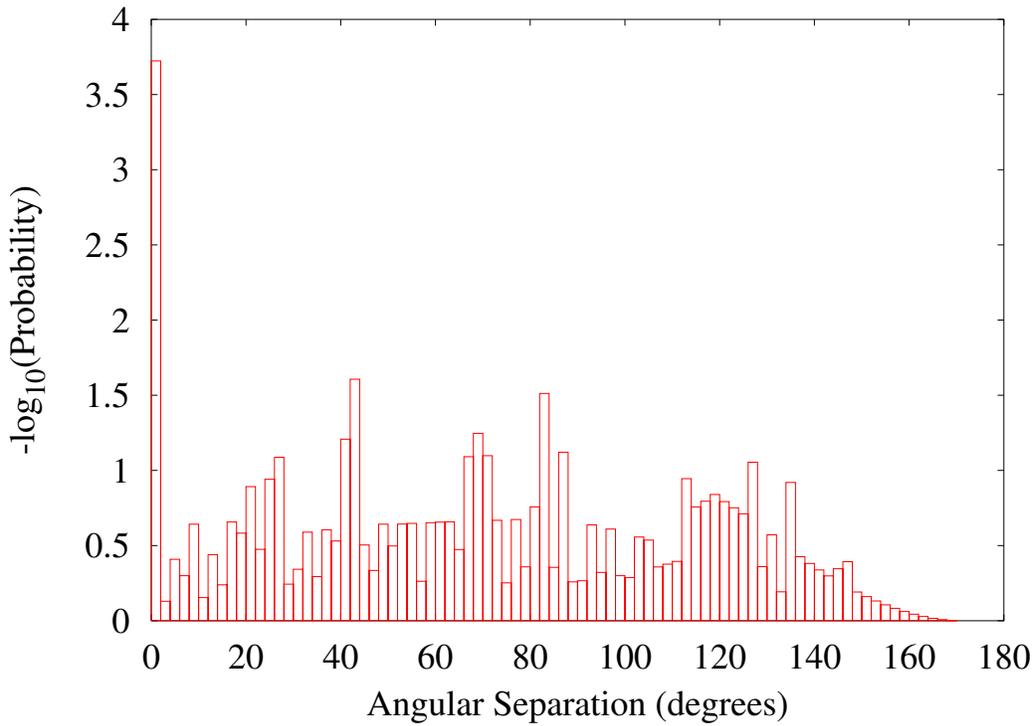}
\caption{Integral Poisson probability of detecting the observed number of coincident pairs, given 
the mean value as determined by the simulation.  For separations greater then 4 degrees the number of 
coincident pairs is consistent with a uniform distribution of hot-spots.  For small angular separations 
there exists a statistically significant excess number of correlations.}
\end{figure}

\clearpage
\begin{deluxetable}{clccccc}
\tabletypesize{\scriptsize}
\tablecaption{Details of the surveys done by the Milagro, Tibet 2001, and Tibet 2003.}
\tablewidth{0pt}
\tablehead{\colhead{Obs.} &\colhead{Ang. Resolution} &\colhead{Dates of Exposure} &\colhead{Dec. Region (deg.)} &\colhead{N $\geq$4 $\sigma$} &\colhead{Threshold Energy (TeV)}}
\startdata
Milagro  & 0.75 & Jan. 2001 to Dec. 2003 & 1.1 to 80 & 11 & 0.2$^{a}$\\
Tibet 2001 &0.9 &Feb. 1997 to Oct. 1999 & 10 to 50 & 19 & 3$^{b}$ \\
Tibet 2003 &0.9 &Nov. 1999 to June 2001 & 0 to 60 & 23 & 3$^{b}$
\enddata
\tablenotetext{a}{Milagro reports to be sensitive to gamma rays above 200 GeV and reports a median energy of 4 TeV \citep{milagro_crab}.  
In the Atkins et. al. 2004 the median energy of Milagro is shown as a function of declination and spectral index.}
\tablenotetext{b}{Tibet reports the mode of the energy distribution and the reported angular resolutions 
are for energies greater then the mode.}
\end{deluxetable}

\clearpage
\begin{deluxetable}{cccrrrrl}
\tabletypesize{\scriptsize}
\tablecaption{Co-located hot-spots from Milagro\citep{milagro_all_sky}, Tibet 2001\citep{tibet_all_sky_2001} and Tibet 2003\citep{tibet_all_sky_2003}.  The last column shows the upper limits determined by the Milagro group.  The Tibet 2001 and the Tibet 2003 analyses did not report upper limits.}
\tablewidth{0pt}
\tablehead{
\colhead{Pair}& \colhead{No.} &\colhead{Survey} & \colhead{RA} & \colhead{Dec} & \colhead{$\sigma$} & \colhead{Flux Limits(Crab Flux)}
}
\startdata
A &1 & Milagro$^{a}$  &  306.6  & 38.9 & 4.2  & 0.78\\
B & 2 & Milagro$^{a}$  &  313.0  & 32.2 & 4.5  & 0.85\\
C & 3 & Milagro$^{a}$  &  356.4  & 29.5 & 4.1  & 0.84\\
A & 4 & Tibet 2003$^{b}$& 304.15 & 36.45& 4.0  & NA\\
A & 5 & Tibet 2001$^{c}$& 305.4  & 37.9 & 4.15 & NA\\
B & 6 & Tibet 2001$^{c}$& 313.5  & 32.4 & 4.27 & NA\\
C & 7 & Tibet 2001$^{c}$& 358.0  & 30.1 & 4.10 & NA 
\enddata
\tablecomments{Excesses corresponding to known source locations have been excluded(Crab and Mrk 421)}

\tablenotetext{a}{Total number of excesses above 4$\sigma$ is 9}
\tablenotetext{b}{Total number of excesses above 4$\sigma$ is 21}
\tablenotetext{c}{Total number of excesses above 4$\sigma$ is 18}
\end{deluxetable}

\clearpage

\begin{deluxetable}{ccc}
\tabletypesize{\scriptsize}
\tablecaption{Calculated chance probability of having exactly 
{\it N} coincident hot-spot pairs using two different methods}
\tablewidth{0pt}
\tablehead{
\colhead{{\it N}} &\colhead{Method 1} &\colhead{Method 2}
}
\startdata
0 & 94.5\% & 96.1\%\\
1 & 5.4\%  & 3.7\%\\
2 & 0.1\%  & 0.16\%\\
3 & 0.003\%& 0.011\%
\enddata
\end{deluxetable}

\end{document}